\documentclass[
aps,
prd,
showpacs,
twocolumn,
]{revtex4-2}

\usepackage{amsmath}
\usepackage{multirow}
\usepackage{graphicx}
\usepackage{enumerate}
\usepackage[english]{babel}
\usepackage{xcolor}

\begin{document}


\title{Partial wave analysis of $\psi(3686)\rightarrow K^+K^-\eta$}




\author{
\begin{small}
\begin{center}
M.~Ablikim$^{1}$, M.~N.~Achasov$^{10,d}$, P.~Adlarson$^{59}$, S. ~Ahmed$^{15}$, M.~Albrecht$^{4}$, M.~Alekseev$^{58A,58C}$, A.~Amoroso$^{58A,58C}$, F.~F.~An$^{1}$, Q.~An$^{55,43}$, Y.~Bai$^{42}$
% ...
\\
\vspace{0.2cm}
(BESIII Collaboration)\\
\vspace{0.2cm} {\it
$^{1}$ Institute of High Energy Physics, Beijing 100049, People's Republic of China\\
% ...
$^{4}$ Bochum Ruhr-University, D-44780 Bochum, Germany\\
% ...
$^{10}$ G.I. Budker Institute of Nuclear Physics SB RAS (BINP), Novosibirsk 630090, Russia\\
% ...
$^{15}$ Helmholtz Institute Mainz, Johann-Joachim-Becher-Weg 45, D-55099 Mainz, Germany\\
% ...
$^{42}$ Southeast University, Nanjing 211100, People's Republic of China\\
$^{43}$ State Key Laboratory of Particle Detection and Electronics, Beijing 100049, Hefei 230026, People's Republic of China\\
% ...
$^{55}$ University of Science and Technology of China, Hefei 230026, People's Republic of China\\
% ...
$^{58}$ (A)University of Turin, I-10125, Turin, Italy; (B)University of Eastern Piedmont, I-15121, Alessandria, Italy; (C)INFN, I-10125, Turin, Italy\\
$^{59}$ Uppsala University, Box 516, SE-75120 Uppsala, Sweden\\
% ...
$^{d}$ Also at the Novosibirsk State University, Novosibirsk, 630090, Russia\\
% ...
}\end{center}

\vspace{0.4cm}
\end{small}

}

\date{\today}

\begin{abstract}
Using a sample of $(448.1\pm2.9)\times10^6$ $\psi(3686)$ events collected 
with the BESIII detector, we perform the first partial wave analysis 
of $\psi(3686)\rightarrow K^+K^-\eta$.  
In addition to the well established states, $\phi(1020)$, $\phi(1680)$, 
and $K_3^*(1780)$, contributions from $X(1750)$, $\rho(2150)$, $\rho_3(2250)$, 
and $K^*_2(1980)$ are also observed.  
The $X(1750)$ state is determined to be a $1^{--}$ resonance.  
The simultaneous observation of the $\phi(1680)$ and $X(1750)$ indicates 
that the $X(1750)$, with previous observations in photoproduction, 
is distinct from the $\phi(1680)$.  The masses, widths, branching 
fractions of $\psi(3686)\rightarrow K^+K^-\eta$, and the intermediate resonances are also measured.
\end{abstract}

\pacs{13.25.Gv, 14.40.Be, 14.40.Df}

\maketitle


\section{Introduction}
Within the framework of the relativistic quark model~\cite{Godfrey:1985xj},
a spectrum similar to that of a heavy quarkonia is expected for
the strangeonium ($s\bar{s}$) sector~\cite{Barnes:2002mu}.
A comprehensive study of the strangeonium spectrum is useful 
to test the theoretical models and also in the search 
for light exotica (resonances that are not dominantly $q\bar{q}$ 
states, often with nonexotic quantum numbers).  
Strangeonia have been studied in different experiments,
such as the study of the initial-state radiation~\cite{Aubert:2006bu, Aubert:2007ur,
Aubert:2007ym, Shen:2009zze, Lees:2011zi},
$J/\psi$ and $\psi(3686)$ decays~\cite{Ablikim:2007ab, Ablikim:2012xy, Ablikim:2014pfc, Ablikim:2019tqd},
 and photoproduction data~\cite{Busenitz:1989gq, Atkinson:1984cs, Aston:1981tb, Link:2002mp}.
However, the strangeonium spectrum 
is much less well understood, and only a few states have been established. Given the unsatisfactory 
knowledge of strangeonium states, a search for missing states predicted by the relativistic quark 
model is necessary to improve the knowledge of the strangeonium spectrum.  
As proposed in Ref.~\cite{Liu:2015zqa}, 
the available high statistics data collected by the BESIII experiment offer excellent opportunities 
to explore the strangeonium spectrum through $J/\psi$ and $\psi(3686)$ decays.

Using $1.06\times 10^8$ $\psi(3686)$ events collected in 2009, BESIII reported a study of
$\psi(3686)\rightarrow K^+K^-\pi^0$ and $\psi(3686)\rightarrow K^+K^-\eta$~\cite{Ablikim:2012xy}.
Two structures are evident in the $K^+K^-$ mass spectrum in $\psi(3686)\rightarrow 
K^+K^-\eta$, and further study of these structures with larger data samples 
is needed.  The BESIII experiment has collected a sample of 
$(448.1\pm2.9)\times10^6$ $\psi(3686)$ 
events~\cite{Ablikim:2017wyh}, about 4 times larger than the sample used in 
Ref.~\cite{Ablikim:2012xy}, which enables such a reexamination.  
In addition, the larger statistics also allows for a study 
of the $K^*$ states in the $K^\pm\eta$ mass spectrum.  
In this paper, we present a partial wave analysis (PWA) of 
$\psi(3686)\rightarrow K^+K^-\eta$, which investigates 
the intermediate states in both mass spectra.

\section{BESIII detector and Monte Carlo simulation}

The BESIII detector is a magnetic
spectrometer~\cite{Ablikim:2009aa} located at the Beijing Electron
Positron Collider (BEPCII)~\cite{Yu:IPAC2016-TUYA01}. The
cylindrical core of the BESIII detector consists of a helium-based
 multilayer drift chamber (MDC), a plastic scintillator time-of-flight
system (TOF), and a CsI(Tl) electromagnetic calorimeter (EMC),
which are all enclosed in a superconducting solenoidal magnet
providing a 1.0~T magnetic field. The solenoid is supported by an
octagonal flux-return yoke with resistive plate counter muon
identifier modules interleaved with steel. The acceptance of
charged particles and photons is 93\% over a $4\pi$ solid angle. The
charged-particle momentum resolution at $1~{\rm GeV}/c$ is
$0.5\%$, and the $dE/dx$ resolution is $6\%$ for the electrons
from Bhabha scattering. The EMC measures photon energies with a
resolution of $2.5\%$ ($5\%$) at $1$~GeV in the barrel (end cap)
region. The time resolution of the TOF barrel part is 68~ps, while
that of the end cap part is 110~ps. 

Simulated samples produced with the {\sc
geant4}-based~\cite{geant4} Monte Carlo (MC) package, which
includes the geometric description of the BESIII detector and the
detector response, are used to determine the detection efficiency
and to estimate the backgrounds. The simulation includes the beam
energy spread and initial state radiation (ISR) in the $e^+e^-$
annihilations modeled with the generator {\sc
kkmc}~\cite{ref:kkmc}. 
The inclusive MC sample consists of the production of the $J/\psi$
resonance, and the continuum processes incorporated in {\sc
kkmc}~\cite{ref:kkmc}.
The known decay modes are modeled with {\sc
evtgen}~\cite{ref:evtgen} using branching fractions taken from the
Particle Data Group~\cite{pdg}, and the remaining unknown decays
from the charmonium states with {\sc
lundcharm}~\cite{ref:lundcharm}. Final state radiation (FSR)
from charged final-state particles is incorporated with the {\sc
photos} package~\cite{photos}.

\section{Event Selection}\label{selection}
Candidate events for $\psi(3686)\rightarrow K^+ K^- \eta, \eta
\rightarrow\gamma\gamma$ are required to have two charged
tracks with opposite charge and at least two photons.
Charged tracks in the polar angle ($\theta$) range
$|\cos\theta|<0.93$ are reconstructed using hits in
the MDC. Charged tracks are required to pass within
$\pm10$ cm of the interaction point (IP) in the direction parallel to the beam 
and within 1~cm of the IP in the plane perpendicular
to the beam. The combined information from the energy loss
($\mathrm{d}E/\mathrm{d}x$) measured in the MDC and the flight
 time in the TOF is used to form particle identification (PID) 
confidence levels for the $\pi$, $K$ and $p$ hypotheses.
A charged track is identified as a kaon if its PID confidence level
for the kaon hypothesis is larger than that for the pion and
proton hypotheses. Both charged tracks for candidate
events are required to be identified as kaons.
Photon candidates are required to have an energy deposit
in the EMC of at least 25 MeV in the barrel ($|\cos\theta|<0.80$)
or 50 MeV in the end caps ($0.86<|\cos\theta|<0.92$).
To eliminate showers from charged particles, photon
candidates must have an opening angle of at least $10^{\circ}$ 
from all charged tracks. 
To suppress electronic noise and showers unrelated to the event,
the EMC time difference from the event start time is required to 
be within [0, 700] ns.

A four-constraint (4C) kinematic
fit is performed under the $K^+K^-\gamma\gamma$ hypothesis,
where the total measured four momentum is constrained to the 
four momentum of the initial $e^+e^-$ system.
For events with more than two photon candidates, the
combination with the smallest $\chi^{2}$ is retained.
To reject possible background contributions with more or fewer photons,
the 4C kinematic fits are also performed under the hypotheses
$K^+K^-\gamma$ and $K^+K^-\gamma\gamma\gamma$. Only events
for which the $\chi^2$ value for the signal hypothesis is less than 30
and also less than the $\chi^2$ values for the background hypotheses
are retained.

\begin{figure*}
\includegraphics[width = 0.32\textwidth]{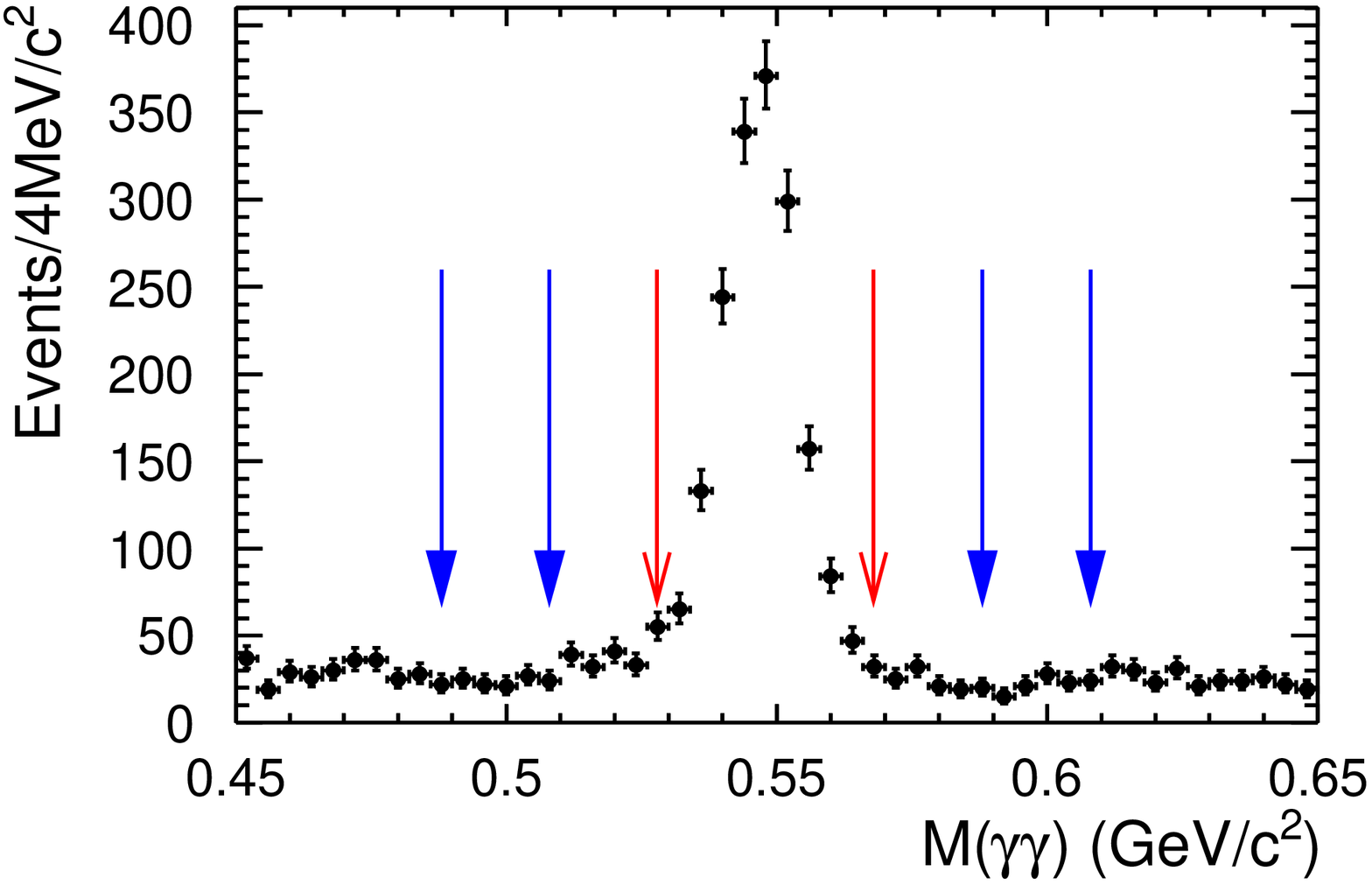}
\put(-95, 80) {(a)}
\includegraphics[width = 0.32\textwidth]{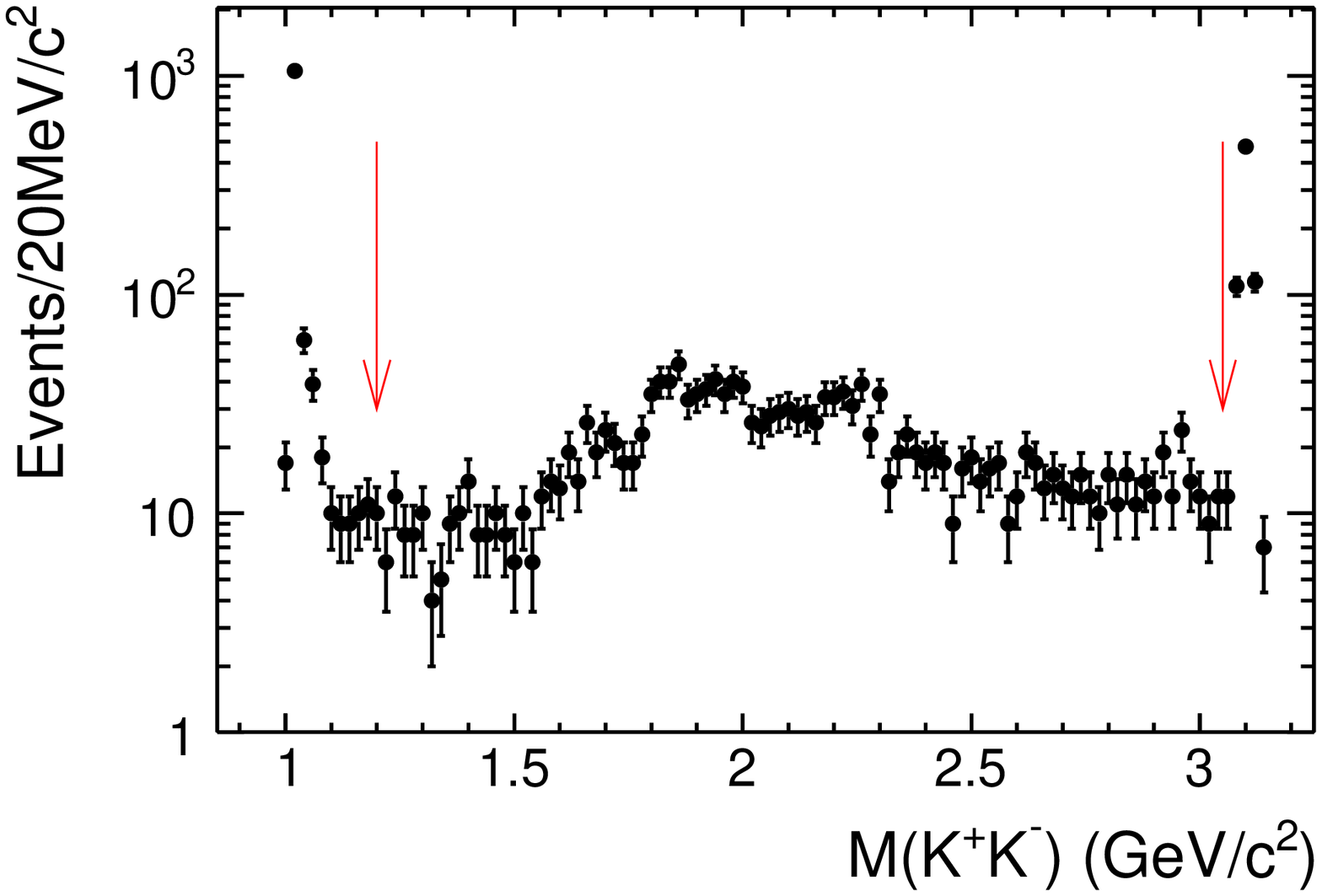}
\put(-95, 80){(b)}
\includegraphics[width = 0.32\textwidth]{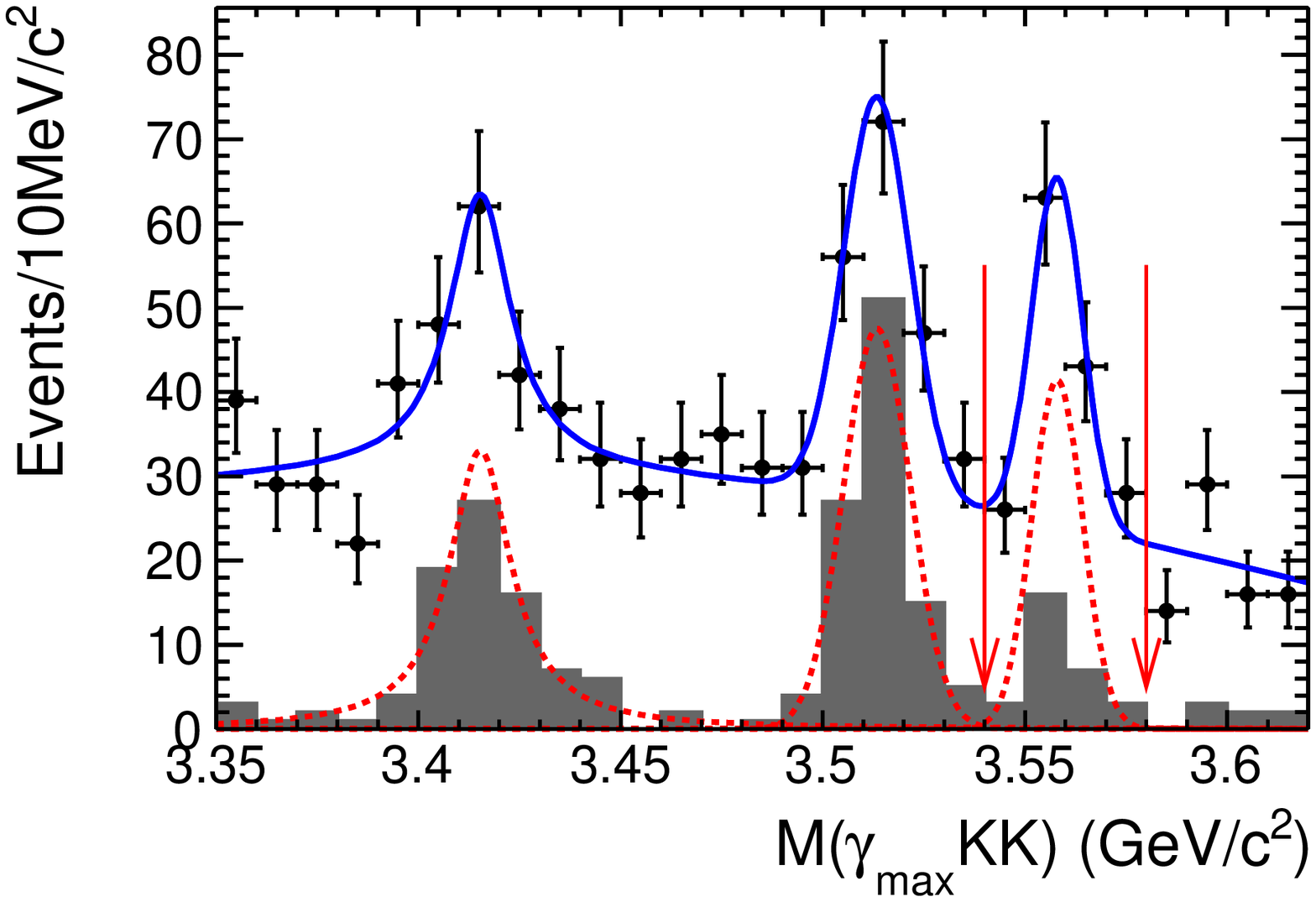}
\put(-95, 80){(c)}
\caption{\label{plot_selection} (a) The $\gamma\gamma$ invariant mass spectrum 
for the data. 
The red arrows show the $\eta$ signal region, while the 
blue arrows with solid arrowheads show the $\eta$ sidebands.
(b) The global $K^+K^-$ invariant mass distribution for the data. Arrows show the 
requirement used to exclude events from $\phi(1020)$ and $J/\psi$ resonances.
(c) The invariant mass distribution of the most energetic photon and two kaons. 
Black markers with error bars show the data in the $\eta$ signal region. 
The grey histograms show the data in the $\eta$ sidebands. The dashed line is
the $\chi_{c0,1,2}$ contribution, from the data in the signal region, which are extracted 
by a global fit (the solid line). Arrows indicate the requirement to exclude the $\chi_{c2}$ events.}
\end{figure*}

The $\gamma\gamma$ invariant mass distribution for events that 
survive the selection criteria is shown in Fig.~\ref{plot_selection}(a), 
where a clear $\eta$ peak is observed. The $K^+K^-$ mass spectrum 
is displayed in Fig.~\ref{plot_selection}(b) after requiring 
$|M(\gamma\gamma) - m_{\eta}|<0.02$ GeV/$c^2$, where $m_{\eta}$ is 
the world average mass of the $\eta$ meson~\cite{pdg}. The two narrow, 
significant peaks correspond to the $\phi(1020)$ and $J/\psi$, respectively,
that come from decays of $\psi(3686)\rightarrow\phi\eta$ and 
$\psi(3686)\rightarrow J/\psi\,\eta$ with the resonances then decaying to $K^+K^-$.  
The $\phi(1020)$ and $J/\psi$ are very well established, and the region between
$m_{\phi(1020)}$ and $m_{J/\psi}$ is more interesting. In this analysis,
only the events in the region 1.20 $< M(K^+K^-) < 3.05$ GeV/$c^2$ are 
used. 

 To investigate possible background contributions, the same analysis is also 
 performed on an inclusive MC sample of $5.06\times 10^8$ $\psi(3686)$ 
 events. The dominant non-$\eta$ background events come from 
$\psi(3686)\rightarrow\gamma\chi_{cJ} (J=0,1,2), \chi_{cJ} \rightarrow K^+K^-\pi^0$ with a missing photon.  
We then investigate the invariant mass of the combination of $K^+K^-$ 
together with the most energetic photon. The $\chi_{cJ}$ peaks are
clearly evident, as indicated in Fig.~\ref{plot_selection}(c), 
where the black markers and grey histograms are data from the 
$\eta$ signal region and sidebands, respectively.  
Unlike the $\chi_{c0,1}$ peaks,
the $\chi_{c2}$ background peak cannot be well estimated with the 
$\eta$ mass sidebands (0.478 $< M(\gamma\gamma) <$ 0.498 GeV/$c^2$
or 0.598 $< M(\gamma\gamma) <$ 0.618 GeV/$c^2$). Therefore, 
the candidate events in the $\chi_{c2}$ mass region of 
3.54 $<M(\gamma_{max}KK) < 3.58$ GeV/$c^2$ are rejected.

\begin{figure}
\includegraphics[width = 0.48\textwidth]{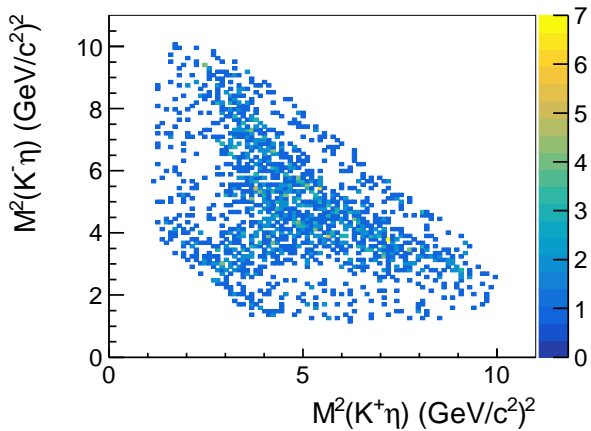}
\caption{\label{plot_dalitz_data} Dalitz plot for selected $\psi(3686)\rightarrow K^+K^-\eta$ events.}
\end{figure}

\begin{figure*}
\includegraphics[width=0.33\textwidth]{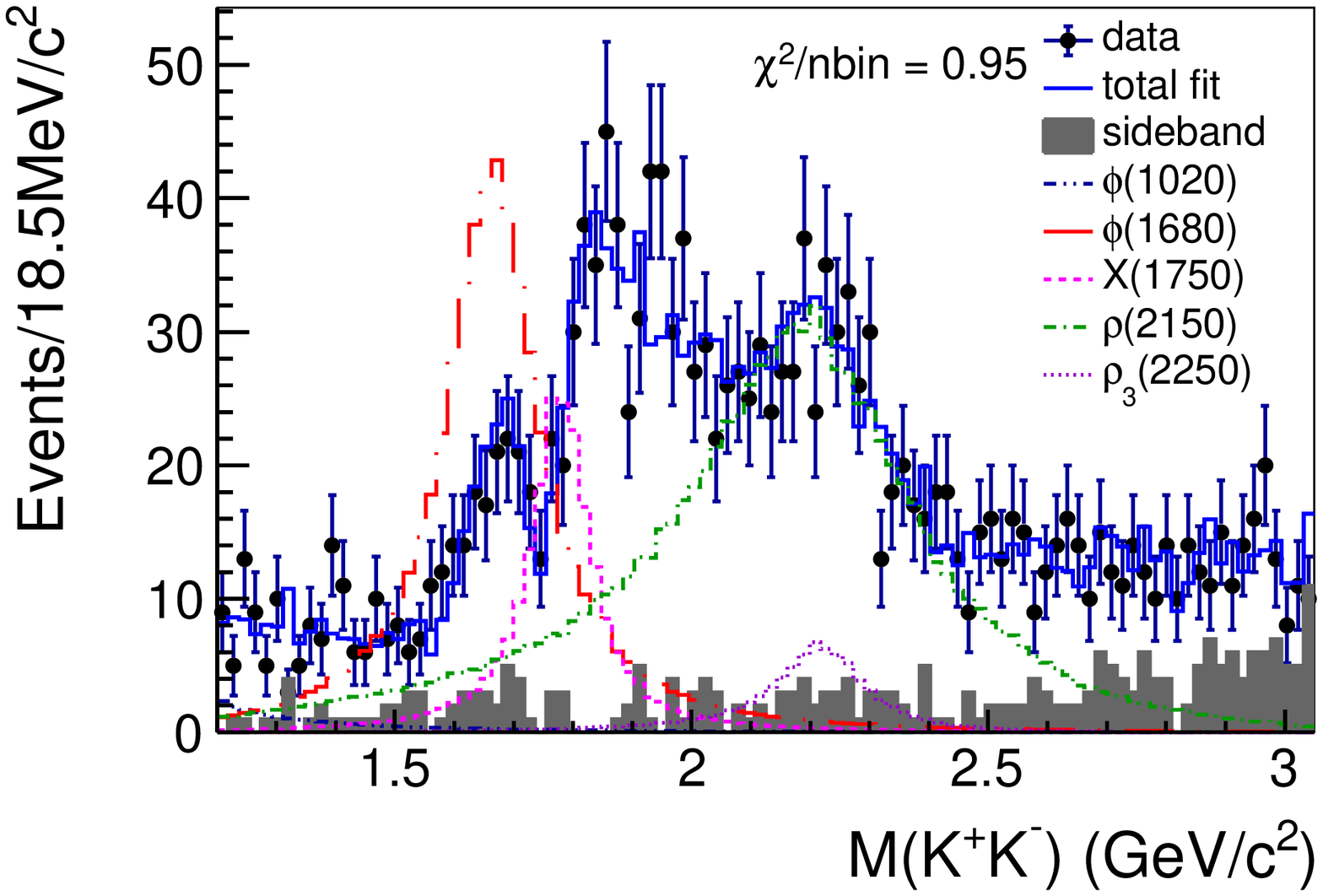} 
\put(-120,90) {(a)}
\includegraphics[width=0.33\textwidth]{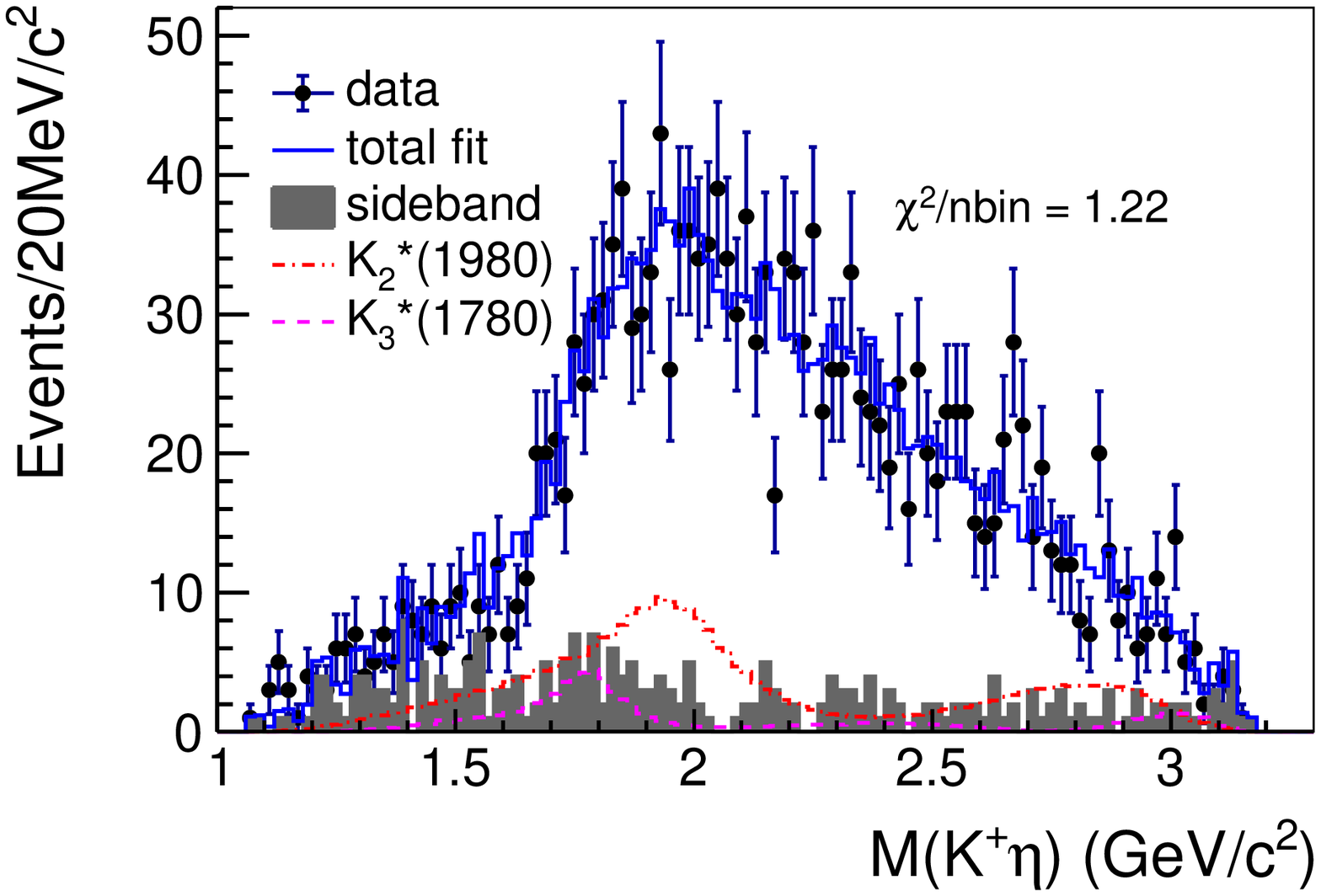} 
\put(-50, 90){(b)}
\includegraphics[width=0.33\textwidth]{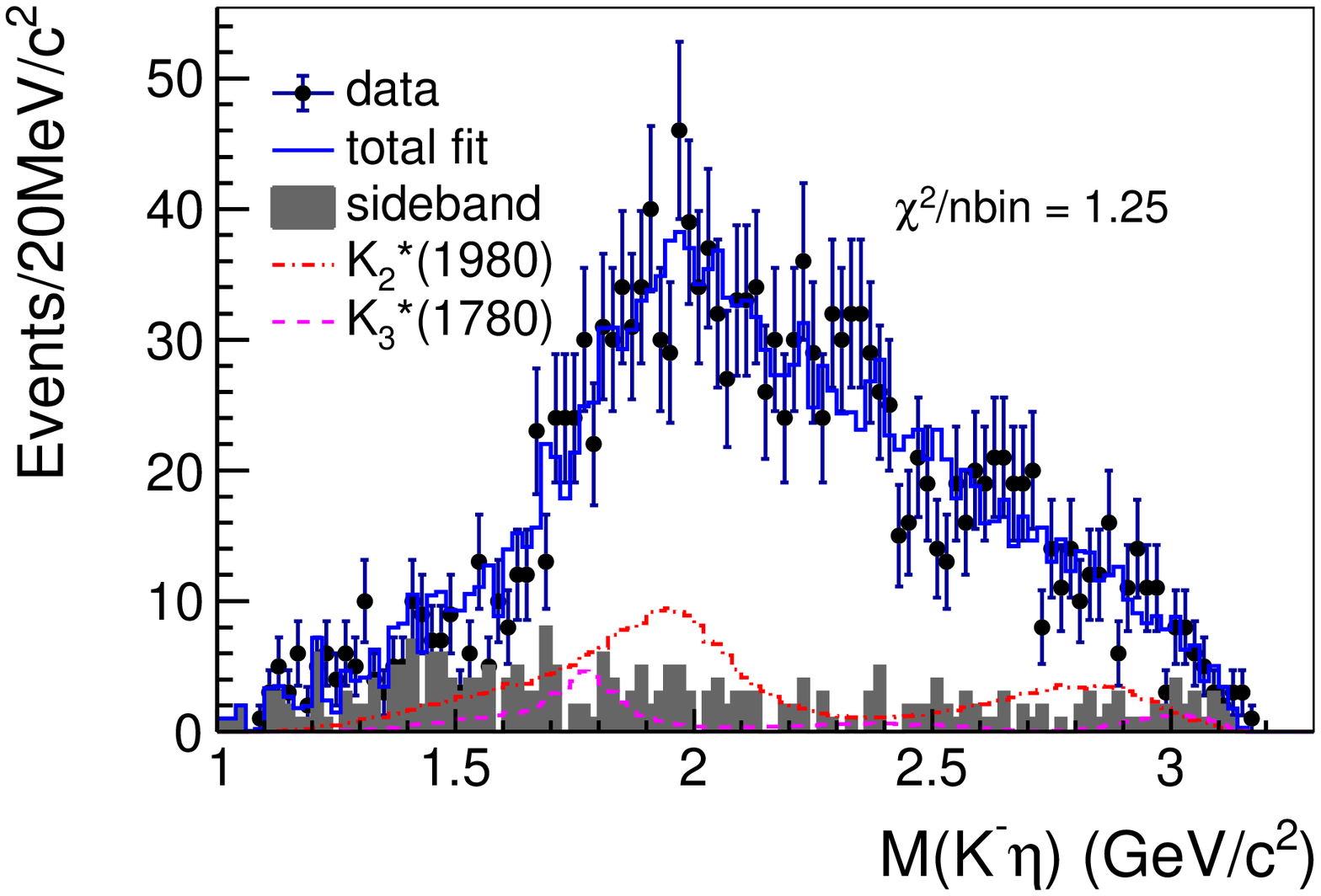} 
\put(-50,90){(c)}
\caption{\label{plot_fit_mass} Comparisons to the fit projections
for the (a) $K^+K^-$, (b) $K^+\eta$, and (c) $K^-\eta$ invariant mass distributions.}
\end{figure*}

After the above requirements, a sample of 1787 $\psi(3686)\rightarrow K^+K^-\eta$ 
 candidates remains. The Dalitz plot for these events, displayed
in Fig.~\ref{plot_dalitz_data}, shows some structures in the distribution.
Structures are also obvious in the $K^+K^-$ mass spectrum shown in Fig.~\ref{plot_fit_mass}(a), 
but not in the $K^+\eta$ and $K^-\eta$ mass spectra shown in Fig.~\ref{plot_fit_mass}(b) 
and Fig.~\ref{plot_fit_mass}(c). Using the $\eta$ mass sidebands, the number of background events is estimated to be 257, as shown by the shaded histograms in 
Figs.~\ref{plot_fit_mass}(a), (b), and (c), and no evident structures are observed in these background $K^+K^-$ and $K^\pm\eta$ mass spectra.

To investigate possible backgrounds from QED processes, which are produced directly in $e^+e^-$
annihilation rather than in $\psi(3686)$ decays, a study is made using a data sample taken 
at $\sqrt{s} =$ 3.773 GeV, with an integrated luminosity of 2.92 fb$^{-1}$~\cite{Ablikim:2014gna}.
After normalizing according to integrated luminosities and the $1/s$ dependence of the cross sections, the background contribution
from QED processes is estimated to be $27.5 \pm 3.1$ events. Due to the low statistics,
this contribution is only considered in the systematic uncertainty due to background contributions.

\section{Partial Wave Analysis}
\subsection{Analysis method}
In the PWA, the decay amplitudes in the sequential decay 
process $\psi(3686)\rightarrow X\eta,X\rightarrow K^+K^-$ and 
$\psi(3686)\rightarrow X^\mp K^\pm, X^\mp\rightarrow K^\mp\eta$ are constructed 
using the covariant tensor formalism described in Ref.~\cite{Zou:2002ar}.
The general form for the decay amplitude is
\begin{equation}
A = \psi_\mu(m)A^\mu = \psi_\mu(m)\sum_i\Lambda_i U_i^\mu,
\end{equation}
where $\psi_\mu(m)$ is the polarization vector of the $\psi(3686)$ and $m$ is
the spin projection of $\psi(3686)$; 
$U^\mu_i$ is the partial wave amplitude with coupling
strength determined by a complex parameter $\Lambda_i$.
The partial wave amplitudes $U_i$ used in the analysis are constructed with the 
four momenta of daughter particles according to the expressions given in Ref.~\cite{Zou:2002ar}.

In this analysis, each intermediate resonance is described by a relativistic Breit-Wigner 
function with an invariant-mass dependent width~\cite{Kuhn:1990ad}
\begin{eqnarray}
	& BW(s) = \frac{1}{m^2 - s - i\sqrt{s}\Gamma(s)}, \\
	& \Gamma(s) = \Gamma_0(m^2)(\frac{m^2}{s})(\frac{p(s)}{p(m^2)})^{2l+1},
\end{eqnarray}
where $s$ is the invariant mass squared of the daughter particles,
$m$ and $\Gamma_0$ are the mass and width of the
intermediate resonance, respectively, $l$ is the orbital angular 
momentum for a daughter particle, and $p(s)$ or $p(m^2)$ is the momentum of 
a daughter particle in the rest frame of the resonance 
with mass $\sqrt{s}$ or $m$.  

The probability to observe the $i$th event characterized by the measurement $\xi_i$, 
i.e., the measured four momenta of the particles in the final state, is
\begin{equation}
P(\xi_i) = \frac{\omega(\xi_i) \, \varepsilon(\xi_i)}{\int\mathrm{d}\Phi \, \omega(\xi) \, \varepsilon(\xi)},
\end{equation}
where $\omega(\xi_i)\equiv(\frac{\mathrm{d}\sigma}{\mathrm{d}\Phi})_i$
is the differential cross section, $\varepsilon(\xi_i)$ is the detection efficiency,
$\mathrm{d}\Phi$ is the standard element of phase space for three-body decays 
and $\int\mathrm{d}\Phi \, \omega(\xi) \, \varepsilon(\xi)=\sigma^\prime$ is
the measured total cross section. The differential cross section is given by \cite{Zou:2002ar} 
\begin{equation}
  \omega = \frac{\mathrm{d}\sigma}{\mathrm{d}\Phi} =
  \frac{1}{2}\sum_{\mu = 1}^2 A^\mu A^{*\mu}, \label{dsigma_dphi}
\end{equation}
where $A^\mu$ is the total amplitude for all possible resonances, 
and $\mu=1,2$ labels the transverse polarization directions.  
Longitudinal polarization is absent since with highly relativistic 
beams $e^+e^-$ annihilation produces $\psi(3686)$ with spin projection 
$J_z = \pm 1$ relative to the beam.  

The likelihood for the data sample is
\begin{equation}
	\mathcal{L} = \prod_{i=1}^N P(\xi_i) = \prod_{i=1}^N
\frac{\omega(\xi_i) \, \varepsilon(\xi_i)}
{\sigma^\prime}.
\end{equation}

Technically, it is more straightforward to minimize negative log-likelihood (NLL), 
$\mathcal{S} = -\ln\mathcal{L}$,
instead of maximizing $\mathcal{L}$, with
\begin{align}\label{seqlnl}
\mathcal{S}  = -\ln\mathcal{L} 
   = -\sum_i^N\ln\left(\frac{\omega(\xi_i)}{\sigma^\prime}\right)
  -\sum_i^N\ln\varepsilon(\xi_i).
\end{align}
In Eq.~\ref{seqlnl}, 
the second term is a constant and has no impact
on the determination of the amplitude parameters or on the relative changes
in $\mathcal{S}$.
In the fit, $-\ln\mathcal{L}$ is defined as
\begin{equation}
  -\ln\mathcal{L} =
    -\sum_i^N\ln\left(\frac{\omega(\xi_i)}{\sigma^\prime}\right)
   = -\sum_i^N\ln\omega(\xi_i)
  + N\ln\sigma^\prime.
\end{equation}

The complex couplings, i.e., the relative magnitudes and phases, of amplitudes 
are determined through an unbinned maximum likelihood fit. The resonance
parameters are optimized by a scan method.  We perform many independent fits 
with varying initial values but with a specific value of the resonance 
parameter under study until a stable minimum negative log-likelihood (MNLL) 
value is obtained.  
We then scan, performing a series of such MNLL searches with various values 
for the resonance parameter; the resonance parameter value with the 
minimum MNLL is taken as our nominal value.  
For each pair of charge conjugate processes and resonances, the two partners 
use the same complex coupling and resonance parameters.

The free parameters in the likelihood function are optimized using 
MINUIT~\cite{James:1975dr}. The measured total cross section $\sigma^\prime$ 
is evaluated using a dedicated MC sample consisting of $N_{\text{gen}}$ events
uniformly distributed in phase space. These events are subjected to the selection 
criteria described in Sec.~\ref{selection} and yield a sample of $N_{\text{acc}}$ accepted events.
The normalization integral is then computed as
\begin{eqnarray}
  \int\mathrm{d}\Phi \, \omega(\xi) \, \varepsilon(\xi) = \sigma^\prime
\rightarrow \frac{1}{N_{\text{gen}}}\sum_{k}^{N_{\text{acc}}}
\omega(\xi_k).
\end{eqnarray}

The background contribution in the fit is estimated using the $\eta$ sideband data and is 
subtracted from the log-likelihood function for data in the $\eta$ signal region, i.e.,
\begin{equation}
	\mathcal{S} = -(\ln\mathcal{L}_{\text{DATA}} - \ln\mathcal{L}_{\text{BG}}).
\end{equation}

The number of the fitted events $N_X$ for an intermediate resonance $X$  
is defined as
\begin{align}
  &N_X = \left( \frac{\sigma_X}{\sigma^\prime} \right) \, N^\prime,  \\
  &\sigma_X = \frac{1}{N_{\text{gen}}}
\sum_{j=1}^{N_{\text{acc}}}\omega_X(\xi_j),
\end{align}
where $N^\prime$ is the number of selected
events after background subtraction and $\omega_X$
denotes the observed differential cross section for the
process with the intermediate state $X$.

The detection efficiency $\varepsilon_X$ for the intermediate
resonance $X$ is obtained using a weighted MC sample that resembles the data,
\begin{equation}
\varepsilon_X = \frac{\sigma_X}{\sigma_X^{\text{gen}}}
   = \frac{\sum_{j=1}^{N_{\text{acc}}}\omega_X(\xi_j)}{\sum_{k=1}^{N_{\text{gen}}}\omega_X(\xi_k)}.
\end{equation}

Taking $\psi(3686)\rightarrow X\eta, X\rightarrow K^+K^-$ as an example, 
the product branching fraction is calculated according to,
\begin{align}
&\mathcal{B}(\psi(3686)\rightarrow X\eta,X \rightarrow K^+K^-) \nonumber = \frac{N_X}{N_{\psi}\cdot\varepsilon_X\cdot\mathcal{B}({\eta\rightarrow\gamma\gamma})}, \\
\end{align}
where $N_\psi$ is the number of $\psi(3686)$
events~\cite{Ablikim:2017wyh} and
$\mathcal{B}(\eta\rightarrow\gamma\gamma)$
is the branching fraction of $\eta\rightarrow\gamma\gamma$~\cite{pdg}.

The free parameters in the fit are the relative magnitudes and phases of the amplitudes.
The statistical uncertainties of the signal yields are propagated from the covariance matrix obtained
from the fit. The statistical uncertainties for the masses and widths, which are optimized using a scan method,
are defined as one standard deviation from the optimized results, corresponding to a change of 0.5
in the log-likelihood value, for a specific parameter.

The statistical significance of a given intermediate resonance is evaluated using
the change in the log-likelihood value and the number of free parameters in the fit with
and without the specific resonance.

\subsection{PWA result}

A PWA is performed on the accepted 1787 candidate events for $\psi(3686)\rightarrow K^+K^-\eta$,
where the background contribution is described with 257 events from the $\eta$ mass sidebands. 
Though most of $\psi(3686)\rightarrow \phi\eta$ events are removed by requiring $M(K^+K^-)>1.2$ GeV/$c^2$,
the amplitude for $\psi(3686)\rightarrow \phi\eta$ is included in the PWA to evaluate its impact on
the interference between the tail of the $\phi$ and other components. However, its contribution is constrained to
the expected number of events, $24.3\pm 2.4$, which is estimated from the branching fraction of 
$\psi(3686)\rightarrow \phi\eta$~\cite{pdg}.

For the other components in the fit, a large number of attempts are made to evaluate the possible 
resonance contributions in the $K^+K^-$ and $K^\pm\eta$ mass spectra~\cite{checked_resonances}. Only components 
with a statistical significance larger than 5$\sigma$ are kept in the baseline solution. In addition to the $\phi$, 
the baseline fit includes contributions from the $\phi(1680)$, $X(1750)$, $\rho(2150)$,  $\rho_3(2250)$, 
$K^*_2(1980)^\pm$, and $K^*_3(1780)^\pm$. The fit results, including the resonance parameters, 
the statistical significance and the product branching fraction for each component, are summarized 
in Table~\ref{tab_sum_res} and Table~\ref{tab_sum_res_bf}. 
Table~\ref{tab_pdg_cand} shows the resonance parameters in baseline solution 
and their average values in Particle Data Group (PDG)~\cite{pdg}.

The spin-parity assignment of the baseline solution is checked for each 
component separately.
Replacing $\phi(1680)$, $\rho(2150)$, or $\rho_3(2250)$ by a $3^{--}$ [$1^{--}$ 
for $\rho_3(2250)$] resonance with same mass and width worsens the NLL values 
by 81.8, 213.8 and 40.1, with the number of degrees of freedom unchanged.  
Altering the $K_2^{*}(1980)$ spin parity to $1^-$, $3^-$, $4^+$ or 
the $K_3^{*}(1780)$ to $1^-$, $2^+$, $4^+$ worsens the NLL values by 
at least 40 units.  
The spin-parity assignment of the $X(1750)$ as $1^{--}$ is significantly better
than the $3^{--}$ hypothesis, with the NLL values improved by 53.4 units.

\begin{figure*}
\includegraphics[width = 0.48\textwidth]{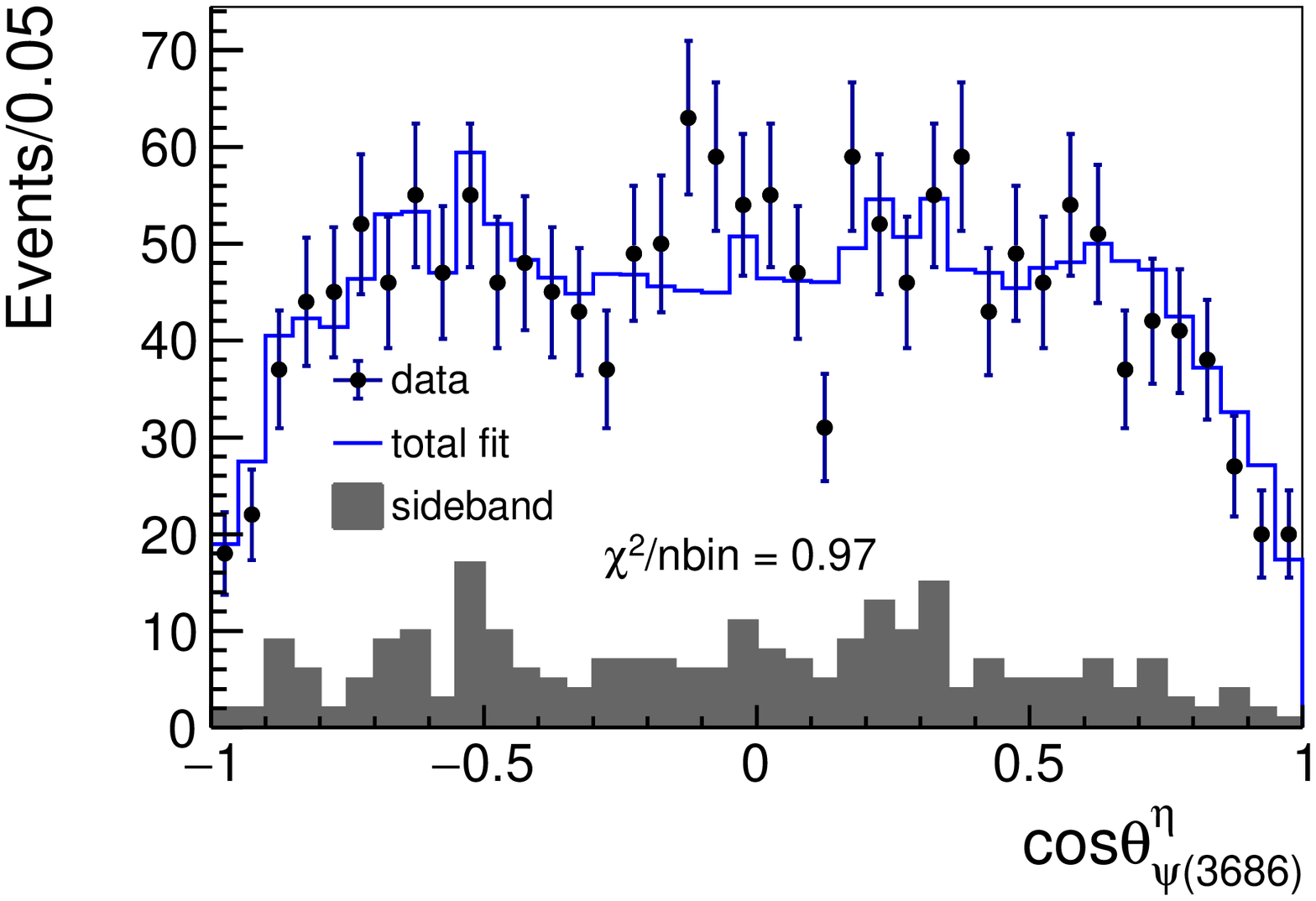} 
\put(-45, 120) {(a)}
\includegraphics[width = 0.48\textwidth]{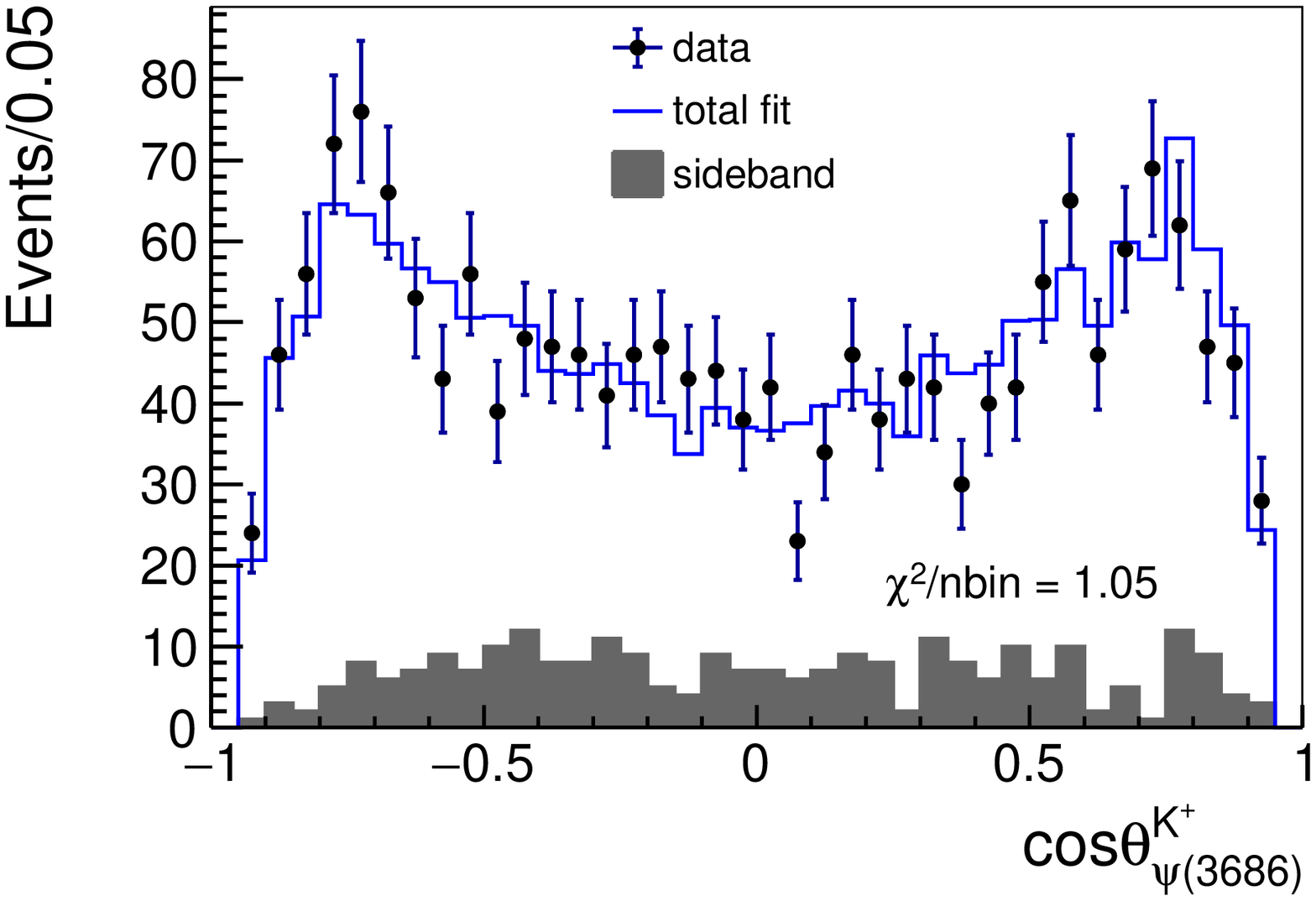} 
\put(-75, 120) {(b)}

\includegraphics[width = 0.48\textwidth]{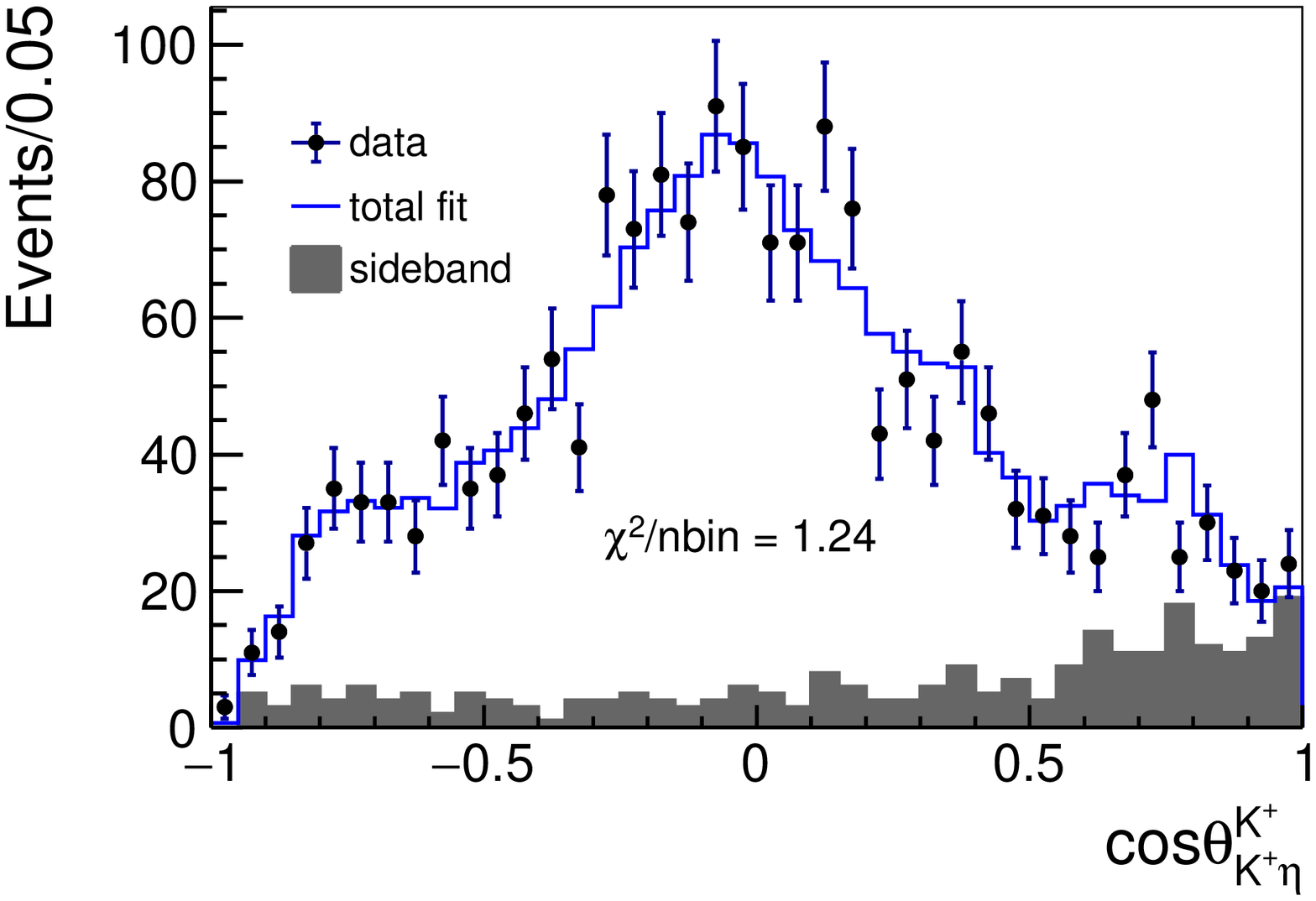} 
\put(-50, 120) {(c)}
\includegraphics[width = 0.48\textwidth]{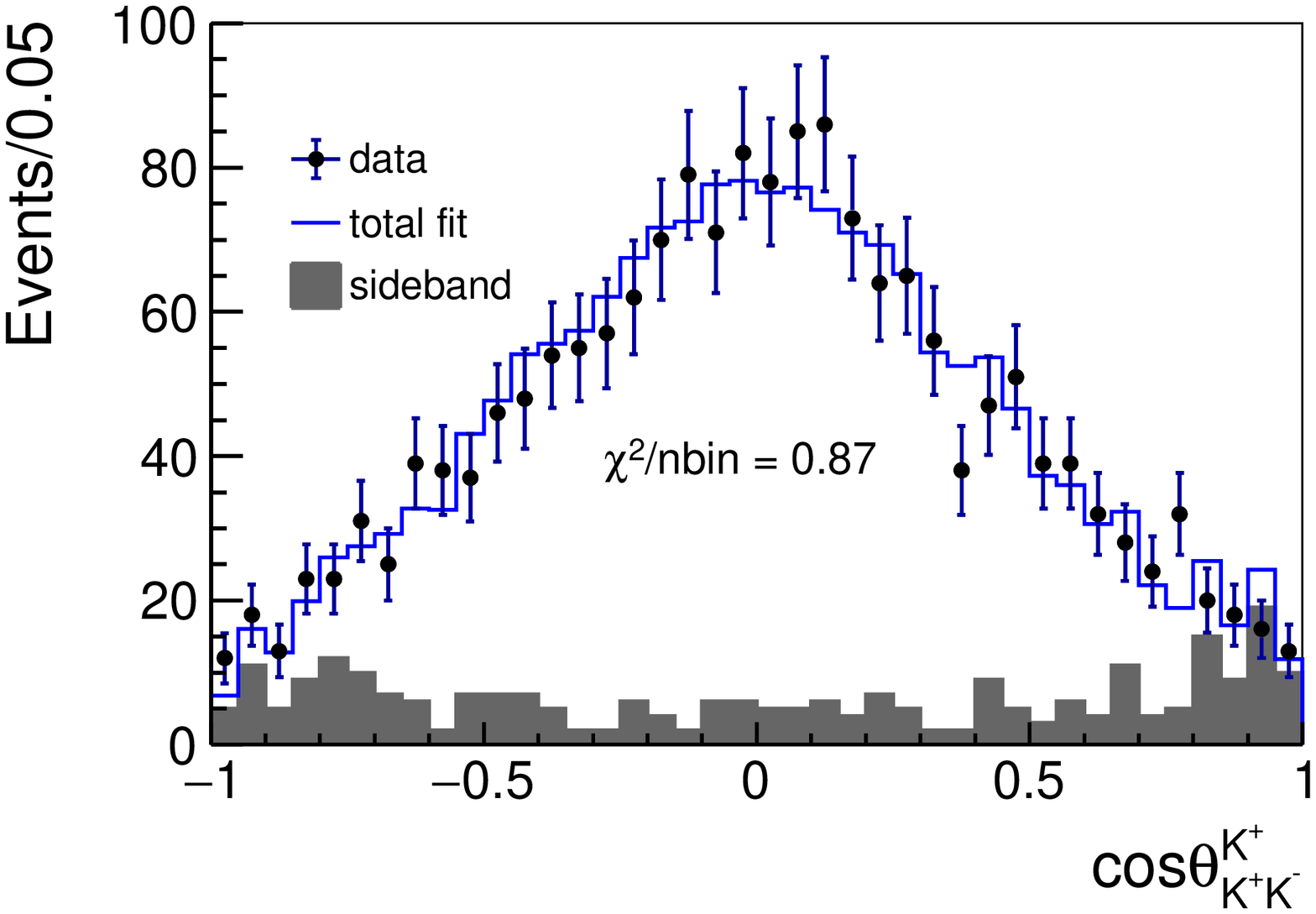} 
\put(-50, 120) {(d)}
\caption{\label{plot_fit_angle} Fit projections to (a) $\cos\theta$ of the $\eta$ in the $\psi(3686)$ rest frame,
(b) $\cos\theta$ of the $K^+$ in the $\psi(3686)$ frame, (c) $\cos\theta$ of the $K^+$ in the $K^+\eta$ rest frame, 
(d) $\cos\theta$ of the $K^+$ in the $K^+K^-$ rest frame.
}
\end{figure*}

The PWA results provide a good description of the data, as illustrated by the comparisons 
between the fit projections and the data for $M(K^+K^-)$, $M(K^+\eta)$, $M(K^-\eta)$, and angular distributions in 
Figs.~\ref{plot_fit_mass} and \ref{plot_fit_angle}. In addition, the comparisons of the angular distributions for the $\eta$ 
($K^\pm$) in different center-of-mass frames also indicate the fit projections are consistent with the data.

In the $K^+K^-$ mass spectrum, the apparent structure around 1.7 GeV/$c^2$ is identified 
in the PWA as the well established $\phi(1680)$. The PWA fit gives a mass of
$1680^{+12+21}_{-13-21}$ MeV/$c^2$ and a width of $185^{+30+25}_{-26-47}$ MeV,
with a statistical significance of 14.3$\sigma$, which are consistent with the world average 
values of the $\phi(1680)$~\cite{pdg}. To describe the clear dip between 1.7 GeV/$c^2$ and 
1.8 GeV/$c^2$, another vector resonant structure, with a statistical significance of $10.0\sigma$, is included in the PWA. Interestingly, the fitted mass 
and width of this structure are $1784^{+12+0}_{-12-27}$ MeV/$c^2$ and 
$106^{+22+8}_{-19-36}$ MeV, respectively, which are in agreement with those of the 
$X(1750)$ reported by the FOCUS Collaboration~\cite{Link:2002mp}. The $X(1750)$ was originally 
interpreted as the photoproduction mode of the $\phi(1680)$ \cite{Busenitz:1989gq,Atkinson:1984cs,Aston:1981tb} 
with the limited statistics. The observation of both the $\phi(1680)$ and the $X(1750)$ in the 
$K^+K^-$ mass spectrum implies that the $X(1750)$ is a new structure instead of the 
photoproduction mode of the $\phi(1680)$. The $\rho(1700)$ is another $1^{--}$ 
resonance in the mass region [1.7, 1.8] GeV/$c^2$. 
The $\rho(1700)$ has a quite different mass and width compared to 
the $X(1750)$, as shown in Table~\ref{tab_pdg_cand}.  
To distinguish the $X(1750)$ and the $\rho(1700)$, an alternative fit is performed after fixing the mass and width of the observed $X(1750)$ 
to instead be those of the $\rho(1700)$~\cite{pdg}. 
This alternate fit yields a likelihood 5.7$\sigma$ worse than the nominal fit.
This test indicates the observed additional vector resonance is more likely to 
be the $X(1750)$ than the $\rho(1700)$. However, the $\rho(1700)$ has a very 
large uncertainty in its mass and width. This large uncertainty of the 
$\rho(1700)$ prohibits excluding the possibility that this vector 
structure is the $\rho(1700)$.

Reference~\cite{Ablikim:2012xy}, which used a subset of the data sample used in this analysis, 
assumed the structure around 2.2 GeV/$c^2$ to be the $\phi(2170)$. By introducing one 
$1^{--}$ component, the PWA fit in this analysis gives $M=2255^{+17+50}_{-18-41}$ 
MeV/$c^2$ and $\Gamma=460^{+54+160}_{-48-90}$ MeV. The width is much larger 
than that of the $\phi(2170)$ from previous measurements~\cite{Aubert:2006bu, Aubert:2007ur, 
Aubert:2007ym, Ablikim:2007ab, Shen:2009zze, Lees:2011zi, Ablikim:2014pfc}.  This structure 
could be either the $\phi(2170)$ or the $\rho(2150)$ or perhaps a superposition of both.
To obtain a good description of the angular distribution, we find that an additional resonance 
with a mass of $2248^{+17+59}_{-17-5}$ MeV/$c^2$ and a width of 
$185^{+31+17}_{-26-103}$ MeV and $J^{PC} = 3^{--}$ is also necessary, which is interpreted as the 
$\rho_3(2250)$ because the mass and width are consistent with previous measurements of
the $\rho_3(2250)$~\cite{Hasan:1994he,Anisovich:2002su}. Due to the low statistics, the 
uncertainties on the resonant parameters for these two structures are quite large. Since, for either excited $\rho$ state, 
$\psi(3686)\rightarrow\rho\eta$ is an isospin violating decay and 
$\rho \rightarrow K^+K^-$ is suppressed by the OZI rule, the investigation of
the $\pi^+\pi^-$ invariant mass in $\psi(3686)\rightarrow\pi^+\pi^-\eta$ may make it possible to 
establish which of these possibilities is correct.

In the $K^\pm\eta$ mass spectra, the fit results indicate that the dominant contributions come 
from the established $K_2^*(1980)$ and $K_3^*(1780)$ mesons. The fitted masses and widths of 
these two resonances, which are summarized in Table~\ref{tab_sum_res}, are consistent with 
their world average values~\cite{pdg}.

\begin{table}
\caption{Mass, width and significance of each component in the baseline solution.
The first uncertainties are statistical and the second are systematic.}
\begin{ruledtabular}
\begin{tabular}{lccc}
Resonance & M (MeV/$c^2$) & $\Gamma$ (MeV)  & Significance \\
\hline
$\phi(1680)$ & $1680^{+12+21}_{-13-21}$ & $185^{+30+25}_{-26-47}$ & $14.3\sigma$ \\
$X(1750)$ & $1784^{+12+0}_{-12-27}$ & $106^{+22+8}_{-19-36}$ & $10.0\sigma$ \\
$\rho(2150)$ & $2255^{+17+50}_{-18-41}$ & $460^{+54+160}_{-48-90}$ & $23.5\sigma$\\
$\rho_3(2250)$ & $2248^{+17+59}_{-17-5}$ & $185^{+31+17}_{-26-103}$ & $8.5\sigma$ \\
$K^*_2(1980)$ & $2046^{+17+67}_{-16-15}$ & $408^{+38+72}_{-34-44}$  & $19.9\sigma$ \\
$K^*_3(1780)$ & $1813^{+15+65}_{-15-16}$ & $191^{+43+3}_{-37-81}$ & $11.2\sigma$ \\
\end{tabular}
\end{ruledtabular}
\label{tab_sum_res}
\end{table}

\begin{table}
\caption{Branching fraction for each process in the baseline solution. 
The first uncertainties are statistical and the second are systematic.}
\begin{ruledtabular}
\begin{tabular}{lc}
Decay mode & BF ($\times10^{-6})$\\
\hline
$\psi(3686)\rightarrow\phi(1680)\eta\rightarrow K^+K^-\eta$  & $12.0\pm1.3^{+6.5}_{-6.9}$ \\
$\psi(3686)\rightarrow X(1750)\eta\rightarrow K^+K^-\eta$  & $4.8\pm1.0^{+2.6}_{-2.6}$ \\
$\psi(3686)\rightarrow\rho(2150)\eta\rightarrow K^+K^-\eta$  & $21.7\pm1.9^{+7.7}_{-8.3}$\\
$\psi(3686)\rightarrow\rho_3(2250)\eta\rightarrow K^+K^-\eta$ & $1.9\pm0.4^{+0.5}_{-1.3}$\\
$\psi(3686)\rightarrow K^*_2(1980)^\pm K^\mp  \rightarrow K^+K^-\eta$  & $7.0\pm0.5^{+3.7}_{-0.6}$\\
$\psi(3686)\rightarrow K^*_3(1780)^\pm K^\mp  \rightarrow K^+K^-\eta$  & $2.0\pm0.4^{+1.9}_{-0.4}$\\
\end{tabular}
\end{ruledtabular}
\label{tab_sum_res_bf}
\end{table}

\begin{table*}
\caption{Comparison of resonances parameters in the baseline solution and 
their average values in PDG. The first uncertainties are statistical and the second are systematic.}
\begin{ruledtabular}
\begin{tabular}{lcccc}
\multirow{2}{*}{Resonance}  
    & \multicolumn{2}{c}{This work} & \multicolumn{2}{c}{PDG~\cite{pdg}} \\
    & M (MeV/$c^2$) & $\Gamma$ (MeV)  & M (MeV/$c^2$) & $\Gamma$ (MeV) \\
\hline
$\phi(1680)$ & $1680^{+12+21}_{-13-21}$ & $185^{+30+25}_{-26-47}$ & $1680\pm20$ & $150\pm50$ \\
\multirow{2}{*}{$X(1750)$} & \multirow{2}{*}{$1784^{+12+0}_{-12-27}$} & \multirow{2}{*}{$106^{+22+8}_{-19-36}$}
     & $(1720\pm20)_{\rho(1700)}$ & $(250\pm100)_{\rho(1700)}$ \\
  &  &  
     & $(1753.5\pm1.5\pm2.3)_{X(1750)}$~\cite{Link:2002mp} & $(122.2\pm6.2\pm8.0)_{X(1750)}$~\cite{Link:2002mp}\\
\multirow{2}{*}{$\rho(2150)$} & \multirow{2}{*}{$2255^{+17+50}_{-18-41}$} & \multirow{2}{*}{$460^{+54+160}_{-48-90}$}
     & $(2153\pm27)_{\rho(2150)}$~\cite{Biagini:1990ze} & $(389\pm79)_{\rho(2150)}$~\cite{Biagini:1990ze} \\
  &  &
     & $(2175\pm15)_{\phi(2170)}$ & $(61\pm18)_{\phi(2170)}$ \\
$\rho_3(2250)$ & $2248^{+17+59}_{-17-5}$ & $185^{+31+17}_{-26-103}$ & $2232$~\cite{Hasan:1994he} & $220$~\cite{Hasan:1994he}\\
$K^*_2(1980)$ & $2046^{+17+67}_{-16-15}$ & $408^{+38+72}_{-34-44}$  & $1973\pm8\pm25$ & $373\pm33\pm60$ \\
$K^*_3(1780)$ & $1813^{+15+65}_{-15-16}$ & $191^{+43+3}_{-37-81}$ & $1776\pm6$ & $159\pm21$\\
\end{tabular}
\end{ruledtabular}
\label{tab_pdg_cand}
\end{table*}

\section{Branching fraction of $\psi(3686)\rightarrow K^+K^-\eta$}
The comparisons of different mass spectra and angular distributions, as displayed in Figs. ~\ref{plot_fit_mass} and ~\ref{plot_fit_angle}, indicate that
the PWA results are in good agreement with the data. In this case, the PWA results provide a good model to simulate 
the decay $\psi(3686)\rightarrow K^+K^-\eta$  and allow a determination of its branching fraction with 
\begin{align}
  & \mathcal{B}(\psi(3686)\rightarrow K^+K^-\eta) \\
  & = \frac{N_{{\rm data}} - N_{{\rm sd}} - N_{\phi\eta} - N_{{\rm QED}}}
  {N_{\psi}\mathcal{B}(\eta\rightarrow\gamma\gamma) \varepsilon} \\
  & = (3.49\pm 0.09 \pm 0.15)\times 10^{-5}, \nonumber 
\end{align}
where $N_{{\rm data}}=1787$ is the number of   $\psi(3686)\rightarrow K^+K^-\eta$ candidates
after excluding $\psi(3686)\rightarrow\phi\eta$ and $\psi(3686)\rightarrow J/\psi\eta$ 
processes with a requirement $1.20 < M(K^+K^-) < 3.05$ GeV/$c^2$. 
The background contribution estimated by $\eta$ sidebands is $N_{{\rm sd}} = 257$.
Contributions from the remaining $\psi(3686)\rightarrow\phi\eta$ and QED processes are 
estimated to be $N_{\phi\eta} = 24.3\pm2.4$ and $N_{{\rm QED}} = 27.5\pm3.1$.
The detection efficiency is determined to be $\varepsilon = 23.95\%$ modeled by the PWA results above.
The first uncertainty is statistical and the second is systematic, which will be discussed below.

\section{Systematic uncertainties}
The systematic uncertainties in the intermediate resonance measurements 
are divided into two categories.
The uncertainties in the first category are applicable to
all branching fraction measurements.
These uncertainties include the systematic uncertainties from
photon detection (1\% per photon~\cite{syst_photon}), MDC tracking (1\%
per charged track~\cite{syst_tracking}), PID (1\% per kaon~\cite{syst_pid}),
number of $\psi(3686)$ events~\cite{Ablikim:2017wyh},
the branching fraction of $\eta\rightarrow\gamma\gamma$
(0.5\% ~\cite{pdg}) and the kinematic fit (1.4\%). The systematic
uncertainty associated with the kinematic fit comes from
the inconsistency of the track-helix parameters between the
data and MC simulation. This difference can be reduced by
correcting the helix parameters of charged tracks in the MC
simulation as described in Ref.~\cite{syst_helix_corr}. The uncertainty
due to the kinematic fit is estimated to be 1.4\% by comparing
the detection efficiency with and without the correction.

The uncertainties in the second category are due to the PWA fit
procedure and are applicable to measurements of both
branching fractions of intermediate states and the corresponding  resonance parameters. Sources of
these uncertainties include impact from the tail of the $\phi(1020)$ resonance,
resonance parametrization, resonance parameters,
background estimation ($\chi_{c2}$ veto, contribution
from QED processes, and sideband region), additional
resonances, and the radius of the centrifugal barrier.
These uncertainties are discussed below.

\begin{enumerate}[(i)]

\item
$\chi_{c2}$ veto:
In the nominal fit, events within the window 3.54 $<M(\gamma_{max}KK) < 3.58$ GeV/$c^2$
are removed. To estimate the uncertainty due
to this requirement, these events are included in the fit, and 
a MC sample of 
$\psi(3686)\rightarrow\gamma\chi_{c2},\chi_{c2}\rightarrow K^+K^-\pi^0$ is used
to describe $\chi_{c2}$ background in the fit.
The MC events are generated in accordance with the amplitude analysis results 
in Ref.~\cite{BESIII:2016dda} to provide a good description of the data.
The differences in the PWA fit results due to this change are taken as
uncertainties. The change of the branching fraction of $\psi(3686)\rightarrow K^+K^-\eta$, 1.1\%,  
with and without this requirement is assigned as the uncertainty from this source.

\item
\label{alt_side_band}
Sideband region:
The events in the $\eta$ sideband region 
(0.478 $<M(\gamma\gamma) < 0.498$ GeV/$c^2$ or
0.598 $<M(\gamma\gamma) < 0.618$ GeV/$c^2$) are used to estimate
the background contribution in the PWA fit. An alternative sideband
region (0.488 $< M(\gamma\gamma) < 0.508$ GeV/$c^2$ or
0.588 $<M(\gamma\gamma)<0.608$ GeV/$c^2$) is also used and the
differences in the fit results relative to the nominal ones are taken as the
associated uncertainties.

\item
The tail of the $\phi(1020)$ resonance:
The $\phi(1020)$ resonance is very narrow and is far away from 
the PWA region. Impacts from the tail of the $\phi(1020)$,
the resolution effect on the $\phi(1020)$ tail and the uncertainty 
of the branching fraction $\psi(3686)\rightarrow\phi(1020)\eta$, are negligible.
As a test, we artificially increase the width of the $\phi(1020)$ 
to 6.27 MeV [$\sim 1.5\times\Gamma(\phi)$] and refit data. The difference 
between this result and the nominal result is found to be negligible.
We also vary the branching fraction of $\psi(3686)\rightarrow\phi(1020)\eta$
by $\pm 1\sigma$ around the world average value in the fit 
and comparing these fit results with the nominal result.
Differences due to variations of the branching fraction
are found to be negligible.

\item
Resonance parametrization:
To estimate the uncertainty due to the resonance parametrization of the 
resonance shape, we performed the PWA by replacing 
the nominal parametrization with
a relativistic Breit-Wigner with a constant width~\cite{Zou:2002ar}
$f = \frac{1}{m^2 - s - im\Gamma}$, where $m$ and $\Gamma$
are the mass and width of the resonance, and $s$ is the
invariant mass squared of the daughter particles.
The differences due to the resonance parametrization are
taken as the systematic uncertainties.

\item
Resonance parameters:
The uncertainty due to resonance parameters
(mass and width) is estimated by varying the parameters
by $\pm1\sigma$ around the nominal results in
the fit, one at a time. The largest changes
after these variations are taken as the systematic
uncertainties.

\item
QED contribution:
The estimated contribution from QED processes is $27.5\pm3.1$ events,
which are not included in the nominal fit. The
uncertainty is estimated by subtracting this contribution
using a datalike MC sample, which includes $K^{*}_2(1430)^{+}K^{-} + c.c.$,
$K^{*}_3(1780)^{+}K^{-} + c.c.$ and $1^{--}$ nonresonant processes.
The MC sample is generated according to a preliminary PWA fit to the 2.92 fb$^{-1}$ data
sample taken at 3.773 GeV~\cite{Ablikim:2014gna}.
The differences between the nominal fit and the fit with
the QED contribution subtracted are taken as systematic
uncertainties.

\item
Additional resonances:
To estimate uncertainties due to additional resonances,
fits with additional resonances are performed. The 
spin 1 resonances~{} $K^{*}(1410)$ ($4.3\sigma$), 
$K^{*}(1680)$ ($3.9\sigma$), and spin 3 resonance 
$\rho_3(1990)$ ($2.2\sigma$)~\cite{pdg} are
included separately. The differences relative to the nominal result
are taken as systematic uncertainties.

\item
Radius of the centrifugal barrier:
The Blatt-Weisskopf barrier factor~\cite{barrier_factor1, barrier_factor2}
is included in the PWA
decay amplitudes and the radius ($R$) of the centrifugal
barrier is used in the factor via $Q_0 = (0.197321/R[\mathrm{fm}])$ 
GeV/$c$~\cite{Zou:2002ar}. In the nominal fit $Q_0$ is set to
0.2708 GeV/$c$. Fits with alternative radii
($Q_0$ = 0.15 GeV/$c$, and $Q_0$ = 0.5 GeV/$c$) are also performed and
the differences relative to the nominal fit result are taken as
systematic uncertainties.

\end{enumerate}

Systematic uncertainties for masses, widths and
branching fractions and the sources described above
are summarized in Table~\ref{sys_mw} and Table~\ref{sys_bf}.
Assuming all the above uncertainties are independent,
the total systematic uncertainty is calculated by adding
them in quadrature.

\begin{table*}
\caption{Sources of systematic uncertainties and their corresponding contributions  to the mass (in MeV/$c^2$) and width (in MeV) of intermediate resonances.}
\begin{ruledtabular}
\begin{tabular}{lcccccccccccc}
\multirow{2}{*}{Sources} & \multicolumn{2}{c}{$\phi(1680)$} & \multicolumn{2}{c}{$X(1750)$} & \multicolumn{2}{c}{$\rho(2150)$} & \multicolumn{2}{c}{$\rho_3(2250)$} & \multicolumn{2}{c}{$K_2^*(1980)^\pm$} & \multicolumn{2}{c}{$K_3^*(1780)^\pm$} \\\cline{2-3}\cline{4-5}\cline{6-7}\cline{8-9}\cline{10-11}\cline{12-13} 
  &  $\Delta$M & $\Delta\Gamma$ & $\Delta$M & $\Delta\Gamma$ & $\Delta$M & $\Delta\Gamma$ & $\Delta$M & $\Delta\Gamma$ & $\Delta$M & $\Delta\Gamma$ & $\Delta$M & $\Delta\Gamma$ \\
\hline
Breit-Wigner parametrization & $+14.4$ & $-14.3$ & $+0.3$ & $-15.3$ & $+34.9$ & $+70.9$ & $+39.8$ & $+2.5$ & $+54.1$ & $+67.7$ & $+54.1$ & $+0.2$\\
Resonance parameter & $^{+6.0}_{-15.7}$ & $^{+13.4}_{-22.0}$ & $-9.1$ & $-22.0$ & $^{+8.2}_{-8.1}$ & $^{+13.0}_{-21.9}$ & $+13.6$ & $^{+2.7}_{-49.8}$ & $^{+9.2}_{-10.4}$ & $^{+20.1}_{-20.9}$ & $^{+6.2}_{-10.4}$ & $^{+1.1}_{-40.3}$ \\
$\chi_{c2}$ veto & $-3.2$ & $+1.3$ & $-9.9$ & $-2.3$ & $-2.9$ & $+5.2$ & $+24.9$ & $-62.6$ & $+12.6$ & $+11.7$ & $-0.9$ & $-20.9$ \\
Background estimation & $-10.3$ & $-6.0$ & $-3.7$ & $-18.5$ & $-14.7$ & $+131.2$ & $+1.7$ & $-51.7$ & $-7.7$ & $-31.5$ & $+5.5$ & $-43.4$\\
Continuum background & $-3.7$ & $-1.6$ & $-6.9$ & $-8.3$ & $+1.0$ & $-4.4$ & $+9.4$ & $-24.3$ & $+1.5$ & $+1.7$ & $-0.7$ & $-7.2$\\
Additional resonances & $+14.5$ & $^{+11.0}_{-33.2}$ & $-20.3$ & $^{+8.1}_{-4.6}$ & $^{+25.2}_{-28.2}$ & $^{+43.9}_{-48.9}$ & $+25.9$ & $^{+16.1}_{-15.8}$ & $^{+33.3}_{-6.7}$ & $-18.8$ & $^{+33.4}_{-6.7}$ & $^{+2.2}_{-42.1}$\\
Barrier Radius& $-6.8$ & $^{+17.9}_{-18.5}$ & $-9.5$ & $-11.2$ & $^{+24.7}_{-24.1}$ & $^{+32.6}_{-71.9}$ & $^{+19.1}_{-5.0}$ & $-27.1$ & $+16.3$ & $^{+4.4}_{-13.4}$ & $^{+12.2}_{-9.8}$ & $^{+0.3}_{-28.0}$ \\
\hline
Total & $^{+21.3}_{-20.6}$ & $^{+25.0}_{-46.6}$ & $^{+0.3}_{-27.3}$ & $^{+8.1}_{-35.8}$ & $^{+50.3}_{-40.8}$ & $^{+159.5}_{-89.8}$ & $^{+59.3}_{-5.0}$ & $^{+16.5}_{-103.2}$ & $^{+67.4}_{-14.6}$ & $^{+71.7}_{-44.3}$ & $^{+65.3}_{-15.8}$ & $^{+2.5}_{-80.9}$\\
\end{tabular}
\end{ruledtabular}
\label{sys_mw}
\end{table*}

\begin{table*}
\caption{Sources of systematic uncertainties and their corresponding contributions (in \%) to the branching fraction for each decay process.}
\begin{ruledtabular}
\begin{tabular}{lccccccc}
Sources & $\phi(1680)$ & $X(1750)$ & $\rho(2150)$ & $\rho_3(2250)$ & $K_2^*(1980)^\pm$ & $K_3^*(1780)^\pm$ & $\psi(3686)\rightarrow K^+K^-\eta$ \\
\hline
Photon detection & 2.0 & 2.0 & 2.0 & 2.0 & 2.0 & 2.0 & 2.0\\
MDC tracking & 2.0 & 2.0 & 2.0 & 2.0 & 2.0 & 2.0 & 2.0\\
Particle ID & 2.0 & 2.0 & 2.0 & 2.0 & 2.0 &2.0 &2.0 \\
Kinematic fit & 1.4 & 1.4 & 1.4 & 1.4 & 1.4 & 1.4 &1.4 \\
$\mathcal{B}(\eta\rightarrow\gamma\gamma)$ & 0.5 & 0.5 & 0.5 & 0.5 & 0.5 & 0.5 & 0.5\\
Number of $\psi(3686)$ events & 0.6 & 0.6 & 0.6 & 0.6 & 0.6 & 0.6 & 0.6 \\
$\chi_{c2}$ veto & $-18.2$ & $-20.2$ & $-8.8$ & $+5.7$ & $-5.9$ & $-14.1$ & 1.1\\
Background estimation & $-31.7$ & $-3.5$ & $-27.9$ & $-1.4$ & $+25.3$ & $+7.0$ & 1.4\\

Breit-Wigner parametrization & $+2.8$ & $+8.5$ & $-8.6$ & $+6.1$ & $+14.0$ & $+5.2$ & -\\
Resonance parameter & $^{+53.7}_{-30.5}$ & $^{+49.4}_{-35.8}$ & $^{+10.9}_{-7.8}$ & $^{+12.7}_{-14.5}$ & $^{+8.5}_{-3.0}$ & $^{+8.9}_{-6.8}$ & -\\

Continuum background & $-5.7$ & $+7.9$ & $-0.9$ & $+6.6$ & $+0.2$ & $+6.5$ & -\\
Additional resonances & $-29.3$ & $-36.2$ & $^{+16.6}_{-14.7}$ & $^{+18.5}_{-60.4}$ & $+41.9$ & $^{+93.0}_{-9.1}$ & -\\
Barrier radius& $^{+5.3}_{-12.2}$ & $+21.5$ & $^{+29.4}_{-15.8}$ & $^{+11.6}_{-28.9}$ & $^{+7.1}_{-5.0}$ & $+11.8$ & -\\
\hline
Total & $^{+54.2}_{-57.6}$ & $^{+55.2}_{-55.0}$ & $^{+35.7}_{-38.4}$ & $^{+27.7}_{-68.6}$ & $^{+52.2}_{-9.1}$ & $^{+94.8}_{-18.5}$ & $4.2$
\end{tabular}
\end{ruledtabular}
\label{sys_bf}
\end{table*}

\section{Summary}
Using a sample of $4.48\times10^{8}$ $\psi(3686)$ events collected with the BESIII 
detector, we perform a partial wave analysis of $\psi(3686)\rightarrow K^+K^-\eta$ for the first time.  
After excluding contributions from $\psi(3686)\rightarrow\phi\eta$ and $\psi(3686)\rightarrow J/\psi\eta$ 
processes, the branching fraction of $\mathcal{B}(\psi(3686)\rightarrow K^+K^-\eta)$ is calculated 
to be $(3.49\pm 0.09\pm 0.15)\times 10^{-5}$.
With the advantage of the higher statistics data set and the precision MC model,
this result is in agreement with but more precision than the previous measurement~\cite{Ablikim:2012xy}.
This measurement supersedes that in Ref.~\cite{Ablikim:2012xy} which 
was based on a subsample of the data used in this work.

In the $K^+K^-$ mass spectrum, in addition to the established $\phi(1680)$,  a $1^{--}$ state is 
necessary to describe the dip around 1.75 GeV/$c^2$, which is caused by the interference between 
the two states. The fitted mass and width of the $1^{--}$ resonance are consistent with those of the $X(1750)$ 
reported by the FOCUS Collaboration~\cite{Link:2002mp}. However, due to the 
large uncertainty in the mass and width of the $\rho(1700)$, the possibility 
that this $1^{--}$ resonance is the $\rho(1700)$ cannot be excluded.
The broad structure around 2.2 GeV/$c^2$
is caused by contributions from a broad $1^{--}$ structure and a $3^{--}$ structure. The likely candidate
for the former state is either the $\phi(2170)$, $\rho(2150)$, or a superposition of both, while the latter 
state may be attributed to the $\rho_3(2250)$. However, it is still difficult to distinguish these states 
from the excited $\phi$ and $\rho$ states due to the limited statistics. With the help of other 
decays, e.g., $\psi(3686)\rightarrow\pi^+\pi^-\eta$, a combined partial wave analysis may help to distinguish 
these states as strangeonium or excited $\rho$ states.

In the $K^\pm\eta$ mass spectra, no clear peak is observed. The partial wave analysis finds that 
the dominant $K^*$ contributions are from two known states, the $K_2^*(1980)$ and $K_3^*(1780)$.

\begin{acknowledgments}
  The BESIII collaboration thanks the staff of BEPCII and the IHEP computing center for their strong support. This work is supported in part by National Key Basic Research Program of China under Contract No. 2015CB856700; National Natural Science Foundation of China (NSFC) under Contracts Nos. 11521505, 11625523, 11635010, 11675184, 11735014; National Natural Science Foundation of China (NSFC) under Contract No. 11835012; National Key Research and Development Program of China No. 2017YFB0203200; the Chinese Academy of Sciences (CAS) Large-Scale Scientific Facility Program; Joint Large-Scale Scientific Facility Funds of the NSFC and CAS under Contracts Nos. U1532257, U1532258, U1732263, U1832207; CAS Key Research Program of Frontier Sciences under Contracts Nos. QYZDJ-SSW-SLH003, QYZDJ-SSW-SLH040; Chinese Academy of Science Focused Science Grant; National 1000 Talents Program of China; 100 Talents Program of CAS; INPAC and Shanghai Key Laboratory for Particle Physics and Cosmology; German Research Foundation DFG under Contract No. Collaborative Research Center CRC 1044, FOR 2359; Istituto Nazionale di Fisica Nucleare, Italy; Koninklijke Nederlandse Akademie van Wetenschappen (KNAW) under Contract No. 530-4CDP03; Ministry of Development of Turkey under Contract No. DPT2006K-120470; National Science and Technology fund; The Knut and Alice Wallenberg Foundation (Sweden) under Contract No. 2016.0157; The Royal Society, UK under Contract No. DH160214; The Swedish Research Council; U. S. Department of Energy under Contracts Nos. DE-FG02-05ER41374, DE-SC-0010118, DE-SC-0012069; University of Groningen (RuG) and the Helmholtzzentrum fuer Schwerionenforschung GmbH (GSI), Darmstadt

\end{acknowledgments}

\end{document}